\documentclass[12pt]{article} 
\usepackage{ucs} 
\usepackage[utf8x]{inputenc} 
\usepackage[T1]{fontenc} 
\usepackage{amsmath,amssymb,wasysym} 
\usepackage{graphicx} 
\usepackage{color} 
\usepackage[unicode]{hyperref} 
\hypersetup{bookmarksopen=true, bookmarksnumbered=true, pdfstartview={FitH}, pdfborder={0 0 0}, colorlinks=true, linkcolor=red, linktocpage=true, citecolor=blue, urlcolor=cyan, unicode=true, pdftitle={Improved Methods for Hypergraphs}, pdfauthor={DJ, WS}, pdfsubject={Background Field Formalism}, pdfkeywords={BFF, }{Π̌, }{Hypergraphs}}
\usepackage{geometry} 
\usepackage{fancyhdr} 
\usepackage[nottoc,notlof,notlot]{tocbibind} 
\usepackage[titles]{tocloft} 

\usepackage{parskip} 
\DeclareUnicodeCharacter{"0393}{\Gamma}
\DeclareUnicodeCharacter{"0394}{\Delta}
\DeclareUnicodeCharacter{"0398}{\Theta}
\DeclareUnicodeCharacter{"039B}{\Lambda}
\DeclareUnicodeCharacter{"039E}{\Xi}
\DeclareUnicodeCharacter{"03A0}{\Pi}
\DeclareUnicodeCharacter{"03A3}{\Sigma}
\DeclareUnicodeCharacter{"03A5}{\Upsilon}
\DeclareUnicodeCharacter{"03A6}{\Phi}
\DeclareUnicodeCharacter{"03A8}{\Psi}
\DeclareUnicodeCharacter{"03A9}{\Omega}
\DeclareUnicodeCharacter{"03B1}{\alpha}
\DeclareUnicodeCharacter{"03B2}{\beta}
\DeclareUnicodeCharacter{"03B3}{\gamma}
\DeclareUnicodeCharacter{"03B4}{\delta}
\DeclareUnicodeCharacter{"03B5}{\epsilon}
\DeclareUnicodeCharacter{"03B6}{\zeta}
\DeclareUnicodeCharacter{"03B7}{\eta}
\DeclareUnicodeCharacter{"03B8}{\theta}
\DeclareUnicodeCharacter{"03D1}{\vartheta}
\DeclareUnicodeCharacter{"03B9}{\iota}
\DeclareUnicodeCharacter{"03BA}{\kappa}
\DeclareUnicodeCharacter{"03BB}{\lambda}
\DeclareUnicodeCharacter{"03BC}{\mu}
\DeclareUnicodeCharacter{"03BD}{\nu}
\DeclareUnicodeCharacter{"03BE}{\xi}
\DeclareUnicodeCharacter{"03C0}{\pi}
\DeclareUnicodeCharacter{"03C1}{\rho}
\DeclareUnicodeCharacter{"03C3}{\sigma} 
\DeclareUnicodeCharacter{"03C4}{\tau}
\DeclareUnicodeCharacter{"03C5}{\upsilon}
\DeclareUnicodeCharacter{"03C6}{\phi}
\DeclareUnicodeCharacter{"03D5}{\varphi}
\DeclareUnicodeCharacter{"03C7}{\chi}
\DeclareUnicodeCharacter{"03C8}{\psi}
\DeclareUnicodeCharacter{"03C9}{\omega}
\DeclareUnicodeCharacter{"21D0}{\Leftarrow}
\DeclareUnicodeCharacter{"0212B}{\AA}
\DeclareUnicodeCharacter{"00B7}{\cdot}
\DeclareUnicodeCharacter{"266A}{\eighthnote}
\DeclareUnicodeCharacter{"266B}{\twonotes}

\definecolor{title}{rgb}{0.7,0.2,0.5}
\definecolor{abst}{rgb}{0.366,0.366,0.266}
\definecolor{sect}{rgb}{1.0,0.0,0.0}
\definecolor{ssect}{rgb}{0.5,0.5,0.0}
\definecolor{sssect}{rgb}{0.3,0.3,0.3}
\definecolor{appsect}{rgb}{0.0,1.0,0.0}
\definecolor{ref}{rgb}{0.0,0.0,1.0}

\newcommand\sect[1]{{\color{sect}\section{#1}}}
\newcommand\subsect[1] {{\color{ssect}\subsection{#1}}}
\newcommand\subsubsect[1] {{\color{sssect}\subsubsection{#1}}}

\newcommand\references[1] {\color{ref} \begin{thebibliography}{[99]} \color{black} #1 \end{thebibliography}}
\numberwithin{equation}{section} 
\newcommand\equ[1] {\begin{equation}#1\end{equation}}
\newcommand\equn[1] {\begin{equation*}#1\end{equation*}}
\newcommand\eqs[1] {\begin{align}#1\end{align}}
\newcommand\eqsn[1] {\begin{align*}#1\end{align*}}
\newcommand\eqsc[1] {\begin{gather}#1\end{gather}}
\newcommand\eqscn[1] {\begin{gather*}#1\end{gather*}}
\newcommand\fig[2] {\begin{figure}[#1]#2\end{figure}}

\newcommand\arXivid[1] {\href{http://arxiv.org/abs/#1}{\tt arXiv:#1}} 
\newcommand\cmp[3] {{\it Commun.\ Math.\ Phys.\ } \href{http://inspirehep.net/search?ln=en&ln=en&p=find+j+"Commun.Math.Phys.,#1,#3"&of=hb&action_search=Search&sf=&so=d&rm=&rg=25&sc=0}{{\bf #1} (#2) #3}} 
\newcommand\cqg[3] {{\it Class.\ Quant.\ Grav.\ } \href{http://inspirehep.net/search?ln=en&ln=en&p=find+j+"Class.Quant.Grav.,#1,#3"&of=hb&action_search=Search&sf=&so=d&rm=&rg=25&sc=0}{{\bf #1} (#2) #3}}

\newcommand\npb[3] {{\it Nucl.\ Phys.\ }{\bf B#1} (#2) #3}
\newcommand\pr[4] {{\it Phys.\ Rev.\ }{\bf #1#2} (#3) #4} 
\newcommand\pl[4] {{\it Phys.\ Lett.\ }{\bf #1#2} (#3) #4}

\renewcommand\( {\left(}
\renewcommand\) {\right)}
\renewcommand\exp[1] {e^{#1}}
\renewcommand\i {\dot{\iota}}
\newcommand\A {{\cal A}}
\newcommand\D {{\cal D}}
\newcommand\J {{\cal J}}
\renewcommand\S {{\cal S}}
\newcommand\Y {{\cal Y}}
\newcommand\cb {\check{□}}
\newcommand\hb {\widehat{□}}
\newcommand\tr {\text{tr}}

\begin{document}
\pagenumbering{alph}
\title{\color{title}\Huge Improved Methods for Hypergraphs}
\author{DJ, WS}
\author{Dharmesh Jain\footnote{\href{mailto:djain@ic.sunysb.edu}{djain@ic.sunysb.edu}}\, , Warren Siegel\footnote{\href{mailto:siegel@insti.physics.sunysb.edu}{siegel@insti.physics.sunysb.edu}\vskip 0pt \hskip 8pt \href{http://insti.physics.sunysb.edu/\~siegel/plan.html}{http://insti.physics.sunysb.edu/$\sim$siegel/plan.html}}\bigskip\\ \emph{C. N. Yang Institute for Theoretical Physics}\\ \emph{State University of New York, Stony Brook, NY 11794-3840}}
\date{} 
\maketitle
\thispagestyle{fancy}
\rhead{YITP-SB-13-05} 
\lhead{\today}
\begin{abstract}
\normalsize We introduce a new background field method for N=2 superspace. (We treat projective hyperspace, but similar remarks apply for the harmonic case.) In analogy to N=1, background gauge fields are in the real representation, so the lowest-dimension potentials are spinor and the usual non-renormalization theorems are manifest. Another consequence is that the R-coordinates disappear from the effective action.
\end{abstract}

{\footnotesize \tableofcontents}

\newpage
\pagenumbering{arabic}
\cfoot{\thepage}\rhead{}\lhead{}

\sect{Introduction}
The construction of background field formalism for N=2 super-Yang-Mills theory (SYM) in projective hyperspace ($\check{Π}$) \cite{ULMR} is an open problem. Such a formalism is desirable for any (non-)supersymmetric theory as it simplifies (loop) calculations and even intermediate steps respect gauge covariance. A major obstacle in solving this problem for the N=2 case seems to be the lack of knowledge relating the gauge connections to the tropical hyperfield $V$, which describes the SYM multiplet for all practical purposes \cite{DJWS2, DJWS4}.

We note that the closely related \cite{KuzHP, DJWS1} N=2 harmonic superspace ($♫$) \cite{GIKOS-1} doesn't encounter this issue as the hyperfield, $V^{(++)}$ describing the SYM multiplet is itself a connection, $A_{\bar{y}}$. In fact, background field formalism in harmonic superspace has quite a straightforward construction \cite{BBKO}. Although the construction has some subtleties, it has been refined in a series of papers along with relevant calculations \cite{BBIKO, BKMOP, BIP, HarmonicCalculations}.

In this paper, we solve the problem of constructing the background field formalism in projective superspace without the need for knowing the connections explicitly in terms of $V$. This is possible by choosing the background fields to be in a `real' representation ($A_y=0$) and the quantum fields to be in the `analytic' representation ($A_ϑ=0$). This is reminiscent of the quantum-chiral but background-real representation used in N=1 superspace \cite{GSR}. What this does is make the effective action independent of $A_y$ and dependent on background fields (like $A_{ϑ}$) with `dimension' greater than $0$ (since the lowest one is a spinor). Non-existence of $0$-dimension background fields (like $A_y$) is a crucial requirement for the non-renormalization theorems to hold as discussed in \cite{MGWS}. This directly leads to a proof of finiteness beyond 1-loop. (A different approach for proof of finiteness has been discussed in \cite{HSW}.)

The coupling of quantum fields to background fields comes through the former's projective constraint alone, which simplifies the vertex structure a lot. The calculations are also simplified at 1-hoop as most $y$-integrals turn out to be trivial since the background fields have trivial $y$-dependence. This means that the $y$-integration effectively vanishes from the effective action and as expected from the supergraph rules, only one $θ$-integration survives at the end of the calculations. We also work in Fermi-Feynman gauge so there are no IR issues to worry about while evaluating the super-Feynman graphs.

Another important aspect is the ghost structure of the theory in this background gauge. Apart from the expected Faddeev-Popov (fermionic $b,c$) and Nielsen-Kallosh (bosonic $E$) ghosts, we require two more extra ghosts, namely real bosonic $X$ and complex fermionic $R$. This is in contrast to N=1 SYM but very similar to the harmonic treatment of N=2 theory. Heuristically, we can even see that such a field content would give a vanishing $β$-function for N=4. 
Moreover, we will see that the loop contributions of $V$ and extra ghosts have spurious divergences arising due to multiple $δ(y)$'s. These are very similar to the `coinciding harmonic' singularities in the $♫$ case, which manifest themselves at 1-loop level via the subtleties regarding regularization of similar looking determinants. However, in $\check{Π}$ case, we do not encounter such striking similarities. Only the divergences turn out to be similar, leading to a cancellation between the vector hyperfield's contribution and that of the extra ghosts. The finite pieces in the effective action are contributed by these extra ghosts only.

\sect{Construction}
This section is mostly built on the ordinary projective superspace construction of SYM detailed in \cite{DJWS2}. We review it briefly below for the sake of continuity. We also use the 6D notation to simplify some useful identities involving background covariant derivatives and moreover, the results carry over to N=1 6D SYM in a trivial manner with this notation.

\vspace{15pt}
\subsect{Projective Review}
The projective hyperspace comprises of usual spacetime coordinates ($x$), four fermionic ones ($θ$) and a complex coordinate on CP$^1$ ($y$). The full N=2 superspace requires four more fermionic coordinates ($ϑ$) in addition to these projective ones. The super-covariant derivatives corresponding to these extra $ϑ$'s define a projective hyperfield ($Φ$) via the constraint $d_{ϑ}Φ=0$. The algebra of the covariant derivatives will be given below but we note here that in the `real' representation (called `reflective' in \cite{DJWS2} and the one we use extensively in this paper) the $d_ϑ$'s are $y$-dependent. Their anti-commutation relation at different $y$'s is all that we need here:
\equn{\{d_{1ϑ},d_{2ϑ}\}≡\{d_ϑ(y_1),d_ϑ(y_2)\}=(y_1-y_2)d_x≡y_{12}d_x.}

The scalar hypermultiplet is described by an `arctic' hyperfield ($Υ$) that contains only non-negative powers of $y$ and the vector hypermultiplet by a `tropical' $V$, which contains all powers of $y$. To construct the relevant actions, the integration over this internal coordinate is defined to be the usual contour integration, with the contour being a circle around the origin (for our purposes in this paper). So, the projective measure simply reads: $∫dw≡∫dx\,d^4θ∮dy\,$ (with the usual factor of $2π\i$ being suppressed). Now, we are ready to delve into the details of the background field formalism.

\vspace{15pt}
\subsect{Background -- Quantum Splitting}
The gauge covariant derivatives, $∇=d+A$, describing N=2 SYM satisfy the following (anti-) commutation relations (written in 6D notation):
\eqsc{\{∇_{aα},∇_{bβ}\}=-\i C_{ab}∇_{αβ}\,,\\
[∇_{aα},∇_{βγ}]=-ε_{αβγδ}W_a^{δ}\,,\\
\{∇_{aα},W_b^{β}\}=\D_{ab}δ_{α}^{β}-\tfrac{\i}{2}C_{ab}f_{α}^{β}\,,\\
[∇_{αβ},∇^{γδ}]=f_α^{[γ}δ_β^{δ]}\,,\\
[∇_{ϑα},∇_y]=∇_{θα}\,,\quad [∇_{θα},∇_y]=0\,, \label{dyd}}
where the SU(2) index $a=(ϑ,θ)$, $W_{a}^{α}$ and $f_{α}^{β}$ are the field strengths, and $\D_{ab}$ are the triplet of auxiliary scalars. The 4D scalar chiral field strength, $W\sim-\i\(∇_4-\i ∇_5\)$ is related to the spinor field strength via appropriate spinor derivatives. We solve the commutation relation for $∇_ϑ$ by writing $∇_ϑ=\exp{Ω}d_ϑ\exp{-Ω}$, where $Ω$ is an unconstrained complex hyperfield. We can do a background splitting of $Ω$ (similar to N=1 superspace) such that
\equ{∇_ϑ=\exp{Ω_Q}\D_ϑ\exp{-Ω_Q}\,,
}
with $\D_ϑ$ being the background covariant derivative. We can now choose `real' representation for the background derivatives independently such that $\A_y=0⇒\D_y=d_y$. This simplifies the $y$-dependence of the connections:
\eqscn{d_y\A_θ=0\,;\quad d_y\A_ϑ=-\A_θ\\
⇒\A_ϑ=\A_ϑ^{(0)}-y\A_θ\,.}
Since these connections have simple $y$-dependence, the $y$-integrals in the effective action can be trivially done. Moreover, the quantum part of the full covariant derivatives then can be chosen to be in `analytic' representation, $i.e.$, $A_y≠0$ and $A_{ϑ}=0$.

The projective (analytic) constraint on hyperfields `lifts' to $∇_ϑΥ=0$ so we can now define a background projective hyperfield $\hat{Υ}(≡\hat{Υ}_B+\hat{Υ}_Q)$ as $Υ=\exp{Ω_Q}\hat{Υ}$ such that $\D_ϑ\hat{Υ}=0$. Then, the scalar hypermultiplet's action reads:
\equ{\S_{Υ}=-∫dw\,\bar{Υ}Υ=-∫dw\,\bar{\hat{Υ}}\exp{\bar{Ω}_Q}\exp{Ω_Q}\hat{Υ}≡-∫dw\,\bar{\hat{Υ}}\exp{V}\hat{Υ}\,.}

The vector hyperfield $V$'s action looks the same as in the ordinary case; the difference being that the $V$ appearing below is only the quantum piece and is background projective:
\equ{\S_V=\frac{\tr}{g^2}∫dx\,d^8θ∑_{n=2}^{∞}\frac{(-1)^n}{n}\(∏_{i=1}^{n}∫dy_i\)\frac{\(e^{V_1}-1\)⋯\(e^{V_n}-1\)}{y_{12}\,y_{23}⋯y_{n1}}\,.}
We know from \cite{DJWS4} that this action should give an expression for $A_y$ and hence the `analytic' representation for quantum hyperfields is a consistent choice. The background dependence of $V$ comes through the projective constraint and the background covariant derivatives only. The following identities will be useful in showing that and deriving other results in the following sections:
\eqsc{∇_ϑ^4\(\tfrac{1}{2}d_y^2\)∇_ϑ^4=\tfrac{1}{2}\hb\,∇_ϑ^4=\left[\tfrac{1}{2}□-W_{ϑ}^α ∇_{θ,α}+\D_{ϑϑ} d_y+\D_{θϑ}\right]∇_ϑ^4\,,\\
∇_{1ϑ}^4∇_{2ϑ}^4=\left[y_{12}\D_{ϑϑ}+\tfrac{1}{2}y_{12}^2\cb+\tfrac{1}{2}y_{12}^3\(∇_{θ,α}∇^{αβ}∇_{θ,β}+W_{θ}^α ∇_{θ,α}+2\D_{θθ}\)+y_{12}^4∇_{2θ}^4\right]∇_{2ϑ}^4\,,}
where $□=\tfrac{1}{2}∇_{αβ}∇^{αβ}$ is the gauge-covariant d'Alembertian and $\cb=\hb-2\D_{ϑϑ}d_y$. As the quantum connections do not appear explicitly in the calculations, we will drop the usage of curly fonts to denote the background fields (as has been done above) and also the subscript `$_ϑ$' on $W_{ϑ}^{α}$ from now on.

\vspace{15pt}
\subsect{Quantum}
The quantization procedure in the background gauge proceeds similar to the ordinary case. The ordinary derivatives are now background-covariant derivatives so $□$ gets replaced by $\hb$ (or $\cb$) everywhere. Moreover, we need extra ghosts for the theory to be consistent in this formalism as we elaborate further in the following subsections.

\vspace{10pt}
\subsubsect{Scalar and Vector}
The scalar hypermultiplet is background projective but the structure of its action is still the same as in the ordinary case. That means the kinetic operator appearing in the equations of motion is $d_y^2$, $i.e.$, $d_y^2Υ=0$ still holds. So the derivation of the propagator performed in \cite{DJWS2} goes through after employing these changes: $d_{ϑ}→∇_{ϑ}$ and $□→\hb$:
\equ{\langle\bar{Υ}(1)Υ(2)\rangle=\frac{∇_{1ϑ}^4∇_{2ϑ}^4}{y_{12}^3}\frac{1}{\tfrac{1}{2}\hb}δ^8(θ_{12})δ(x_{12})\,.}

The gauge-fixing for the vector hypermultiplet leading to Faddeev-Popov (FP) ghosts is still similar to the ordinary case and we just quote the results with suitable modifications:
\eqs{\S_{gf}&=\,-\frac{\tr}{2α\,g^2}\int dx\,d^8θ\,dy_1\,dy_2\,V_1\left[\frac{y_1}{y_{21}^3}+\frac{y_2}{y_{12}^3}\right]V_2\,;\label{SgfinV}\\
\S_{FP}&=-\tr\int dx\,d^4\theta\,dy\,\left[\bar{b}\,c+\bar{c}\,b+(y\,b+\bar{b})\frac{V}{2}\left(c+\frac{\bar{c}}{y}\right)+...\right].\label{FPaction}}
The propagators for the FP ghosts are similar to the scalar hypermultiplet and will be written down later.

We will always work in Fermi-Feynman gauge ($α=1$) but let us derive the propagator for $V$ with arbitrary $α$ as this technique will be useful later. We first combine the terms quadratic in $V$ from the above equation and the vector hypermultiplet action to get
\eqs{\S_{V}^{(2)}+\S_{gf}^{(2)}=&-\frac{\tr}{2g^2}\int dx\,d^4θ\,dy_1\,dy_2\,V_1\frac{1}{y_{12}^2}\left[1+\frac{1}{α}\(\frac{y_1}{y_{21}}+\frac{y_2}{y_{12}}\)\right]∇_{1ϑ}^4 V_2\nonumber\\
=&-\frac{\tr}{2g^2}\int dx\,d^4θ\,dy_1\,dy_2\,V_1\frac{1}{y_{12}^2}\left[1+\frac{1}{α}\(-1+\frac{y_1+y_2}{2}δ(y_{12})\)\right]y_{12}^2\left(\frac{1}{2}\cb+\cdots\right) V_2\nonumber\\
=&-\frac{\tr}{2g^2}\int dx\,d^4θ\,dy_1\,dy_2\,V_1\left[1+\frac{-1+y_1δ(y_{12})}{α}\right]\left(\frac{1}{2}\cb+\cdots\right) V_2\,.}
Then, we add a generic real source $J$ to the quadratic gauge-fixed vector action:
\eqs{\S_{V-J}=&-\frac{\tr}{g^2}\left\{\int dx\,d^8θ\,dy_{1,2}\,V_1\left[1+\frac{1}{α}\(\frac{y_1}{y_{21}}+\frac{y_2}{y_{12}}\)\right]\frac{1}{2y_{12}^2}V_2-∫dx\,d^8θ\,dy_2\,J_2V_2\right\}\nonumber\\
=&-\frac{\tr}{g^2}\left\{\int dx\,d^4θ\,dy_{1,2}\,V_1\left[1+\frac{-1+y_1δ(y_{12})}{α}\right]\frac{∇_{1ϑ}^4}{2y_{12}^2}V_2-∫dx\,d^4θ\,dy_2\,\J_2V_2\right\}.\label{SVJ}}
Here, $\J$ is now defined to be (background) projective. Now, equation of motion for $V$ reads
\equ{∫dy_1V_1\left[1+\frac{-1+y_1δ(y_{12})}{α}\right]\frac{∇_{2ϑ}^4}{y_{12}^2}=\J_2\,,\label{eomF}}
which we can solve to write $V$ in terms of $\J$. This amounts to inverting the kinetic operator for $V$ as we will see. Assuming the following ansatz for $V$:
\equ{V_{1}=∫dy_0\,\frac{p+q\,δ(y_{01})}{y_{01}^2}\frac{1}{\(\frac{1}{2}\cb\)^2}\J_0∇_{1ϑ}^4\label{FJK}}
and demanding it satisfy (\ref{eomF}), we are led to $p=\frac{(1-α)}{y_0\,y_1}\,\&\,q=\frac{α}{y_0}$ because
\equn{∫dy_1\left[\frac{(1-α)+αy_1δ(y_{01})}{y_0\,y_1}\right]\left[1+\frac{-1+y_1δ(y_{12})}{α}\right]=δ(y_{02})\,.}
 Plugging (\ref{eomF}) and (\ref{FJK}) in the action (\ref{SVJ}), we get
\equn{S_{V-J}=\frac{\tr}{2g^2}∫dx\,d^4θ\,dy_{1,2}\,\J_1∇_{1ϑ}^4\frac{(1-α)+αy_2\,δ(y_{12})}{y_1y_2\,y_{12}^2}\frac{1}{\(\frac{1}{2}\cb\)^2}\J_2\,,}
which leads to the required propagator, first derived (for the ordinary case) in \cite{FGR3}
\equ{\langle V(1)V(2)\rangle=∇_{1ϑ}^4∇_{2ϑ}^4\frac{(1-α)+αy_2\,δ(y_{12})}{y_1y_2\,y_{12}^2}\frac{1}{\(\frac{1}{2}\cb\)^2}δ^8(θ_{12})δ(x_{12})\,.}
This expression simplifies \cite{DJWS2} for $α=1$ to
\equ{\langle V(1)V(2)\rangle=∇_{1ϑ}^4\frac{δ(y_{12})}{y_1}\frac{1}{\frac{1}{2}\cb}δ^8(θ_{12})δ(x_{12})\,,\label{VpropFF}}
as does the quadratic part of the vector action
\equ{\S_{V}^{(2)}=-\frac{\tr}{4g^2}∫dw\,V\(y\cb\)V\,.}

\vspace{10pt}
\subsubsect{1-loop}
In background field gauge, the gauge fixing function leads to additional ghosts apart from the FP ghosts, which contribute to the 1-loop calculations. To see that, consider the effective action $Γ$ defined by the following functional:
\equ{e^{\i Γ}=∫\D V\,\D b\,\D c\,\D f\,e^{\i \(S_{SYM}(V)+S_{FP}(V,b,c)+S_{avg}(f)\)}Δ(V)δ\(f-V\),\label{expG}}
where $Δ(V)$ is found by the normalization condition $∫\D f\,Δ(V) e^{\i S_{avg}(f)}=1$. It gives
\eqs{Δ^{-1}&=∫\D f\,e^{-\i \frac{\tr}{2}∫dx\,d^8θ\,dy_{1,2}f_1\frac{1}{2}\(\frac{y_1}{y_{21}^3}+\frac{y_2}{y_{12}^3}\)f_2}\nonumber\\
&=∫\D f\,e^{-\i \frac{\tr}{2}∫dx_{1,2}\,d^8θ_{1,2}\,dy_{1,2} f_1\Y_{12}f_2}\nonumber\\
&=\frac{1}{\sqrt{\det(\Y_{12})}}\,.}
So (\ref{expG}) simplifies to
\equn{e^{\i Γ}=∫\D V\,\D b\,\D c\,e^{\i \(S_{SYM}(V)+S_{gf}(V)+S_{FP}(V,b,c)\)} \sqrt{\det(\Y_{12})}\,.}
We can rewrite the last factor as
\equ{\frac{1}{\det \Y_{12}}=∫\D ρ\,\D χ\, e^{\i\S_{ρχ}}=∫\D ρ\,\D χ\, e^{\i \tr ∫dw_1 dw_2 ρ_1 \Y_{12} χ_2}}
where $(ρ,χ)$ are unconstrained hyperfields. Proceeding similar to the harmonic case \cite{BBKO}, we redefine $χ → d_y^2χ$ and introduce Nielsen-Kallosh (NK) ghost $E$ to account for the resulting Jacobian. This means the 1-loop contribution for N=2 SYM coupled to matter simplifies to:
\equ{\i Γ=\left\{-\frac{1}{2}\ln\det_{V}\(y\cb\)+\frac{1}{2}\ln\det_{(ρ,χ)}\(\Y_{12}\,d_{y}^2\)\right\}+\left\{\ln\det_{FP}\(d_y^2\)-\frac{1}{2}\ln\det_{NK}\(d_y^2\)\right\}-\frac{1}{2}\ln\det_{Υ}\(d_y^2\).}
For N=4, the scalar hypermultiplet is in adjoint representation and its contribution will cancel the joint FP and NK ghosts contributions. The remaining two terms have spurious divergences due to multiple $δ(y)$'s but their joint contribution has to be finite, which will turn out to be the case as we develop this section further.

To incorporate the effect of $(ρ,χ)$ fields directly in the path integral, we choose to introduce a real scalar $X$ and a complex fermion $R$ as follows:
\equ{\sqrt{\det\(\Y_{12}\,d_{y}^2\)}=∫\D X\,\D R\,\D \bar{R}\,e^{\i\(\S_{X}+\S_{R}\)},}
where
\equ{\S_X=\frac{\tr}{2}∫dw_1 dw_2 X_1 \Y_{12} d_{y_2}^2 X_2\,;\qquad\S_R=\frac{\tr}{2}∫dw_1 dw_2 \bar{R}_1 \Y_{12} d_{y_2}^2 R_2\,.}
So the background field requires 3 Fermionic ghosts $\(b,c,R\)$ and 2 Bosonic ghosts $(E,X)$ and the full quantum action for N=2 SYM coupled to matter reads:
\[S_{N=2}=[S_{SYM}(V)+S_{gf}(V)]+S_{FP}(V,b,c)+S_{NK}(V,E)+S_{XR}(V,X,R)+S_{Υ}(V,Υ).\]

\vspace{10pt}
\subsubsect{Ghosts}
The FP and NK ghosts are background projective hyperfields. The actions for these ghosts look the same as those in the case of non-background gauge. The action for FP ghosts is given in equation (\ref{FPaction}) and that for NK ghost is similar to the scalar hypermultiplet's action. That means their propagators are straightforward generalizations and read
\eqs{\langle \bar{b}(1)c(2)\rangle &=\frac{∇_{1ϑ}^4∇_{2ϑ}^4}{y_{12}^3}\frac{1}{\frac{1}{2}\hb}δ^8(θ_{12})δ(x_{12})\,,\\
\langle \bar{E}(1)E(2)\rangle &=\frac{∇_{1ϑ}^4∇_{2ϑ}^4}{y_{12}^3}\frac{1}{\frac{1}{2}\hb}δ^8(θ_{12})δ(x_{12})\,.}


Now, we focus on the new ingredient of the background field formalism: the eXtRa ghosts. In the same vein as the vector hypermultiplet, we can simplify the actions of these ghosts. Let us just concentrate on the scalar ghost action as the fermionic ghost can be treated similarly:
\eqsn{\S_{X}=&\,-\frac{\tr}{4}∫dx d^8θ∮dy_{1,2}\,X_1\left[\(\frac{y_1}{y_{21}^3}+\frac{y_2}{y_{12}^3}\)d_{y_2}^2\right]X_2\\
=&\,-\frac{\tr}{4}∫dx d^4θ∮dy_{1,2}\,X_1\left[\(\frac{y_1}{y_{21}}+\frac{y_2}{y_{12}}\)\frac{1}{y_{12}^2}\hb\right]X_2\\
=&\,-\frac{\tr}{4}∫dx d^4θ∮dy_{1,2}\,X_1\left[\frac{-1+y_1δ\(y_{12}\)}{y_{12}^2}\hb\right]X_2\,.}

The $X$ propagator can then be derived in a similar way as the vector propagator with arbitrary $α$. Lets add a source term to the action for X ghost:

\eqsn{\S_{X-J}=&-\frac{\tr}{4}∫dx\,d^8θ\,dy_{1,2}\,X_1\left[\(\frac{y_1}{y_{21}^3}+\frac{y_2}{y_{12}^3}\)d_{y_2}^2\right]X_2+\tr∫dx\,d^8θ\,dy_2\,J_2X_2\nonumber\\
=&-\frac{\tr}{4}∫dx\,d^4θ\,dy_{1,2}\,X_1\(\frac{-1+y_1δ(y_{12})}{y_{12}^2}\)\hb X_2+\tr∫dx\,d^4θ\,dy_2\,\J_2X_2\,.}
The equation of motion for $X$ now reads
\equ{∫dy_1X _1\(\frac{-1+y_1δ(y_{12})}{y_{12}^2}\)\hb=2\J_2\,.\label{Xeom}}
Adopting an ansatz for $X$ (similar to what was done for $V$ before),
\equn{X_1=d_{1ϑ}^4∫dy_0\left[p+qδ(y_{01})\right]\frac{1}{\frac{1}{2}\hb^2}2\J_0\,,}
we find that $p=0$ and $q=\frac{1}{y_0}$ satisfy (\ref{Xeom}). Collecting all the results, the action reduces to
\equn{\S_{X-J}=\frac{\tr}{2}∫dx\,d^4θ\,dy_{1,2}\,\J_1∇_{1ϑ}^4\frac{δ(y_{12})}{y_1}\frac{1}{\(\frac{1}{2}\hb\)^2}\J_2\,,}
which leads to the required propagator
\equ{\langle X(1)X(2)\rangle=∇_{1ϑ}^4∇_{2ϑ}^4\frac{δ(y_{12})}{y_1}\frac{1}{\(\frac{1}{2}\hb\)^2}δ^8(θ_{12})δ(x_{12}).\label{Xprop}}
The propagator for the fermionic $R$ ghost has a similar expression.

\newpage
\sect{Calculations}
Given this new construction of the background field formalism for SYM, we can now employ it to calculate contributions to the effective action coming from different hypermultiplets.

\vspace{15pt}
\subsect{Feynman Rules}
The general rules for constructing diagrams in the background field formalism are similar to the ordinary case discussed in \cite{DJWS2}. However, as expected in this formalism, the quantum propagators form the internal lines of the loops and the external lines correspond to the background fields.

The $\cb$ and $\hb$ operators in the propagators need to be expanded around $□_0$ (the connection-independent part of $□$), which will generate the vertices with the vector connection and background fields. For the extra ghosts, we can further simplify the naïve rules by noticing that the vertices have $\frac{1}{y_{12}^2}$-factor and the propagator will generate such a factor in the numerator due to the presence of $∇_{1ϑ}^4∇_{2ϑ}^4$. Thus, we can remove them from the very start and work with the revised propagator and vertex for the purpose of calculating diagrams. Let us now collect all the relevant Feynman rules below.
\eqsn{\text{Scalar propagator:}\quad &\frac{∇_{1ϑ}^4∇_{2ϑ}^4}{y_{12}^3}\frac{δ^8(θ_{12})}{\frac{1}{2}k^2}\\
\text{Vector propagator:}\quad &∇_{1ϑ}^4\frac{δ(y_{12})}{y_1}\frac{δ^8(θ_{12})}{\frac{1}{2}k^2}\\
\text{FP \& NK ghosts propagator:}\quad &\frac{∇_{1ϑ}^4∇_{2ϑ}^4}{y_{12}^3}\frac{δ^8(θ_{12})}{\frac{1}{2}k^2}\\
\text{XR ghosts propagator:}\quad &∇_{1ϑ}^4\frac{δ(y_{12})}{y_1}\frac{δ^8(θ_{12})}{\frac{1}{2}k^2}\\
\text{Scalar, FP \& NK vertex:}\quad &∫d^4θ\,dy\(\hb-□_0\) \qquad \(\text{use} ∫d^4θ∇^{4}_{ϑ}=∫d^8θ\)\\
\text{Vector vertex (background):}\quad & ∫d^4θ\,dy\,y\(\cb-□_0\)\\
\text{Vector vertex (quantum):}\quad & ∫d^8θ\,dy_{1,...,n}\frac{(-1)^n}{y_{12}y_{23}...y_{n1}}\\
\text{XR ghosts vertex:}\quad &∫d^4θ∫dy_{1,2}\,\left[-1+y_1 δ\(y_{12}\)\right]\(\hb-□_0\)}

\vspace{15pt}
\subsect{Examples\label{examples}}
\paragraph{Scalar}
The one-loop contribution from the scalar hypermultiplet to the effective action can not be written in a fully gauge covariant form with a projective measure. Thus, the diagrammatic calculation required to get this contribution (which includes the UV-divergent piece too) is not accessible via the formalism constructed here. We note that such an issue appears in the N=1 background formalism too when the scalar multiplets in complex representation are considered. The calculations cannot be performed covariantly and explicit gauge fields appear in addition to the connections.


\paragraph{Vector} The contribution to one-loop n-point diagrams from vector hypermultiplet running in the loop would be given by the following:
\eqs{Γ_n^{(V)}           \sim&∫d^4k∫d^4θ_{1,...,n}∫dy_{1,...,n}∇_{1ϑ}^4δ^8(θ_{12})\frac{δ(y_{12})}{y_1}\frac{1}{k_1^2}\,y_1\(W^α(1)∇_{θ,α}+...\) ...\nonumber\\
&∇_{nϑ}^4δ^8(θ_{n1})\frac{δ(y_{n1})}{y_n}\frac{1}{k_n^2}\,y_n\(W^α(n)∇_{θ,α}+...\)\nonumber\\
\sim&∫d^4k∫d^8θ_{1,...,n-1}d^4θ_n∫dy_{1,n}δ^8(θ_{12})δ(y_{1n})\frac{1}{k_1^2}\(W^α(1)∇_{θ,α}+...\) ...\nonumber\\
&∇_{nϑ}^4δ^8(θ_{n1})δ(y_{n1})\frac{1}{k_n^2}\(W^α(n)∇_{θ,α}+...\),\label{Vn}
}
where the numerical subscript on $k$ denotes the external momenta dependence. As usual, to kill the extra $δ^8(θ)$-function, at least four $∇_θ$ should be available from the vertices and so $Γ_2^{(V)}=Γ_3^{(V)}=0$. The first non-vanishing contribution is from the 4-point diagram:
\eqs{Γ_4^{(V)}&=\frac{3\,c_A}{2}∫dy_{1,4}δ(y_{14})δ(y_{41}) ∫d^8θ_1d^4θ_4\,\hat{\A}_4\, δ^8(θ_{14})\(∏_{i=1}^{4}\frac{1}{2}W^α(i)∇_{θ,α}\) ∇_{4ϑ}^4δ^8(θ_{41})\nonumber\\
&=\frac{3\,c_A}{32}∮\frac{dy_1}{2ε\,y_1}∫d^8θ_1d^4θ_4\,\hat{\A}_4\,δ^8(θ_{14})\(W^α(1)W^β(2)W^γ(3)W^δ(4) \,ε_{αβγδ}∇_θ^4\)∇_{4ϑ}^4δ^8(θ_{41})\nonumber\\
&=\frac{3\,c_A}{32}∮\frac{dy_{1}}{2ε\,y_1}∫d^4θ\,\hat{\A}_4\,ε_{αβγδ} W^α(1)W^β(2)W^γ(3)W^δ(4)\,,\label{4pdiv}\\
\text{where, }&\hat{\A}_4\sim∫dk\frac{16}{\(k_1^2\)\(k_2^2\)\(k_3^2\)\(k_4^2\)}.\nonumber}
Too many $δ(y)$'s lead to spurious $\frac{1}{ε}$ singularity, similar to `coinciding harmonic' singularities in $♫$. These will cancel when we take into account the $(X,R)$ ghosts.

\paragraph{Extra Ghosts}
Their combined contribution to one-loop n-point diagrams reads:
\eqs{Γ_n^{(X,R)}\sim&-∫d^4k∫d^4θ_{1,...,n}∫d^2y_{1,...,n}∇_{1ϑ}^4δ^8(θ_{12})\frac{δ(y_{1a,2a})}{y_{1a}}\frac{1}{k_1^2}\,\left[\(-1+y_{1a}δ(y_{1a,1b})\)\right]\nonumber\\
&\(W^α(1)∇_{θ,α}+...\) ...\,∇_{nϑ}^4δ^8(θ_{n1}) \frac{δ(y_{nb,1b})}{y_{nb}}\frac{1}{k_n^2}\left[\(-1+y_{na} δ(y_{na,nb})\)\right]\(W^α(n) ∇_{θ,α}+...\)\nonumber\\
\sim&-∫d^4k∫d^4θ_n∫dy_{1a,...,1b}∇_{1bϑ}^4δ^8(θ_{n1})\frac{\(-1+y_{1a}δ(y_{1a,2b})\)}{y_{1a}}\frac{1}{k_1^2}\nonumber\\
&\(W^α(1) ∇_{θ,α}+...\)...\,\frac{\(-1+y_{na}δ(y_{nb,1b})\)}{y_{1b}}\frac{1}{k_n^2}\(W^α(n) ∇_{θ,α}+...\).\label{FXn}
}
Again, the first non-vanishing contribution is from $n=4$ that has the same $δ(y)^2$ singularity structure as the vector in (\ref{4pdiv}) leading to a cancellation, in addition to the following finite part:
\eqs{Γ_4&=-\frac{3\,c_A}{32}∫dy_{1,2,3,4}\(\frac{1}{y_1y_2y_3y_4}-\frac{δ(y_{12})}{y_2y_3y_4}+...\)\,∫d^4θ\,\hat{\A}_4\,ε_{αβγδ} W^α(1)W^β(2)W^γ(3)W^δ(4)\nonumber\\
&=\frac{3\,c_A}{32}∫d^4θ\,\hat{\A}_4\,ε_{αβγδ}\(W^α(1)W^β(2)W^γ(3)W^δ(4)\)|_{y=0}\,.\label{4pFinite}}
The last line follows because only $y$-independent pieces of $W$'s can survive the $y$-integrals. Till here, we have treated $W$'s as fields depending on individual external momenta and eq. (\ref{4pFinite}) is the complete 4-point effective action. Assuming them to be momentum independent, we can further simplify this expression in case of the U(1) gauge group and perform the integral over loop-momentum to get
\[\hat{\A}_4=\frac{16}{24}\frac{1}{\(4π\)^2}\frac{1}{\(W\overline{W}\)^2},\]
where we used the reduction to 4D for $□_0→□_0-2W\overline{W}$. Using this and the fact that $W^α$ is related to $\D_{ϑ}^{α}W \(\&\,\D_{ϑ}^{\dot{α}}\overline{W}\)$, we get the same non-holomorphic 4-point contribution (with the full superspace measure $∫d^8θ$) to N=4 SYM action rather directly when compared to the calculation done in \cite{FGR} (for similar calculations in $♫$ see, for example, \cite{BIP}).

\paragraph{2-loops} We can also see that there are no UV divergences at two-loops. The proof is similar to that given in the ordinary case, \emph{i.e.}, absence of sufficient $∇_{ϑ}^4$'s. Only 3 diagrams shown in fig. \ref{V2Hbg} are supposed to contribute at 2-loops. All of them will vanish due to the $d$-algebra unless we get at least 4 $∇_θ$'s from the expansion of the propagators. This, as we have seen before, brings in 4 more $□$'s making these 2-loop diagrams convergent.
\fig{h}{\centering{\includegraphics[width=8cm]{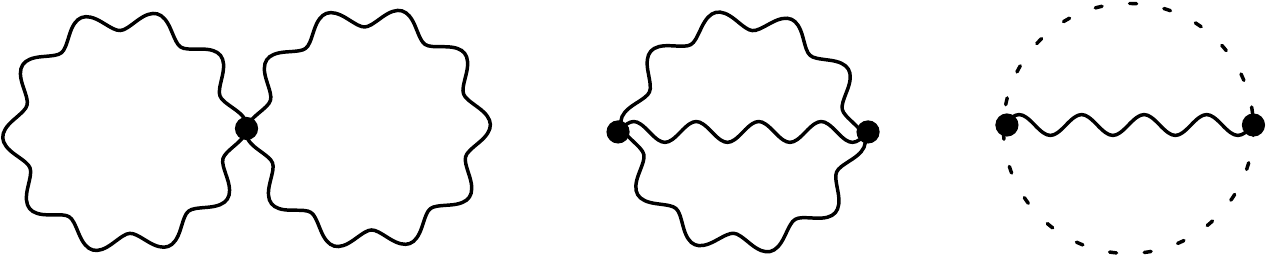}\caption{Diagrams contributing to SYM effective action at 2-loops with external background lines being suppressed.} \label{V2Hbg}}}

Furthermore, we note that the arguments of \cite{MGWS} apply in our case since there is no background connection $A_y$, there cannot be any divergences at 2 or more loops from just power counting. This situation is different than $♫$ where such `0-dimensional' connections are present and arguments similar to the one given above involving number of $∇_ϑ$'s have to be used and at higher loops they can be quite involved \cite{HarmonicCalculations}.

\sect{Conclusion}
We have formulated the background field formalism for N=2, 4D projective superspace. The crucial ingredient was to recognize that different representations for background and quantum pieces of the hypermultiplets are required. Choosing real representation for the background fields allowed non-renormalization theorems to be applicable here as the lowest-dimensional fields available were spinors. The usual choice of analytic representation for the quantum fields allowed us to make a simple extension of the existing `ordinary' super-Feynman rules to the background covariant rules.

Moreover, there are extra ghosts required (apart from FP and NK ghosts) to evaluate the full SYM effective action. These extra ghosts also appear in the harmonic case but in projective case, they cancel the spurious `harmonic' divergences coming from vector hypermultiplet in a straightforward manner and the resultant finite pieces are as expected for N=4. The UV divergent parts come only from the usual (FP and NK) ghosts and scalar hypermultiplet. However, their contribution can not be directly calculated in the formalism developed here for reasons mentioned in section \ref{examples}. We also gave a diagrammatic 2-loops argument for finiteness of N=2 SYM coupled with matter. This is easily supplanted by the power counting argument of \cite{MGWS} in general, which directly leads to a proof for finiteness beyond 1-loop.

For N=1 background formalism, there exist improved rules as showcased in \cite{MGDZ, TRM} and our hope is that such techniques could be applied to what we have developed in this paper. That would lead to a further simplification of the higher-loop calculations while also allowing explicit inclusion of the scalar hypermultiplet's 1-loop contribution.

\vspace{30pt}
\section*{\centering\color{sect}Acknowledgements}
This research work is supported in part by NSF grant no. PHY-0969739.

\references{
\bibitem{ULMR}
U. Lindstr\"{o}m and M. Ro\v{c}ek, \cmp{115}{1988}{21}; \cmp{128}{1990}{191}.
\bibitem{DJWS2}
W. Siegel, 2010, \arXivid{1005.2317} {\color{cyan}\small [hep-th]};\\
D. Jain and W. Siegel, \pr{D}{83}{2011}{105024} [\arXivid{1012.3758} {\color{cyan}\small [hep-th]}];\\
D. Jain and W. Siegel, \pr{D}{86}{2012}{065036} [\arXivid{1106.4601} {\color{cyan}\small [hep-th]}].
\bibitem{DJWS4}
D. Jain and W. Siegel, \pr{D}{86}{2012}{125017} [\arXivid{1203.2929} {\color{cyan}\small [hep-th]}].
\bibitem{KuzHP}
S. M. Kuzenko, \emph{Int. J. Mod. Phys.} {\bf A14} (1999) 1737 [\arXivid{hep-th/9806147}].
\bibitem{DJWS1}
D. Jain and W. Siegel, \pr{D}{80}{2009}{045024} [\arXivid{0903.3588} {\color{cyan}\small [hep-th]}].
\bibitem{GIKOS-1}
A. Galperin, E. Ivanov, S. Kalitzin, V. Ogievetsky and E. Sokatchev, \cqg{1}{1984}{469};\\
E. Ivanov, A. Galperin, V. Ogievetsky and E. Sokatchev, \cqg{2}{1985}{601}; \cqg{2}{1985}{617};\\
A.S. Galperin, E.A. Ivanov, V.I. Ogievetsky, and E.S. Sokatchev, {\it Harmonic superspace} (Cambridge Univ. Press, 2001).
\bibitem{BBKO}
I. L. Buchbinder, E. I. Buchbinder, S. M. Kuzenko and B. A. Ovrut, \pl{B}{417}{1998}{61} [\arXivid{hep-th/9704214}];\\
I. L. Buchbinder and S. M. Kuzenko, \emph{Mod. Phys. Lett.} {\bf A13} (1998) 1623 [\arXivid{hep-th/9804168}].
\bibitem{BBIKO}
I. L. Buchbinder, E. I. Buchbinder, E. A. Ivanov, S. M. Kuzenko and B. A. Ovrut, \pl{B}{412}{1997}{309} [\arXivid{hep-th/9703147}];\\
I. L. Buchbinder, S. M. Kuzenko and B. A. Ovrut, \pl{B}{433}{1998}{335} [\arXivid{hep-th/9710142}].
\bibitem{BKMOP}
E. I. Buchbinder, I. L. Buchbinder and S. M. Kuzenko, \pl{B}{446}{1999}{216} [\arXivid{hep-th/9810239}];\\
S. M. Kuzenko and I. N. McArthur, \pl{B}{506}{2001}{140} [\arXivid{hep-th/0101127}];\\
S. M. Kuzenko and I. N. McArthur, \pl{B}{513}{2001}{213} [\arXivid{hep-th/0105121}].
\bibitem{BIP}
I. L. Buchbinder, E. A. Ivanov and A. Y. Petrov, \npb{653}{2003}{64} [\arXivid{hep-th/0210241}].
\bibitem{HarmonicCalculations}
I. L. Buchbinder and A. Yu. Petrov, \pl{B}{482}{2000}{429} [\arXivid{hep-th/0003265}].
\bibitem{GSR}
M. T. Grisaru, W. Siegel and M. Roček, \npb{159}{1979}{429}.
\bibitem{MGWS}
M. T. Grisaru and W. Siegel, \npb{201}{1982}{292}.
\bibitem{HSW}
P. S. Howe, K. S. Stelle, P. C. West, \pl{B}{124}{1983}{55}.
\bibitem{FGR3}
F. Gonzalez-Rey, 1997, \arXivid{hep-th/9712128}.
\bibitem{FGR}
F. Gonzalez-Rey and M. Roček, \pl{B}{434}{1998}{303} [\arXivid{hep-th/9804010}].
\bibitem{MGDZ}
M. T. Grisaru and D. Zanon, \pl{B}{142}{1984}{359};\\
M. T. Grisaru and D. Zanon, \npb{252}{1985}{578}.
\bibitem{TRM}
T. R. Morris, \pl{B}{164}{1985}{315}.
}
\end{document}